\documentclass[12pt,preprint]{emulateapj}

\usepackage{amsmath,natbib,graphicx}
\usepackage{epsf}
\usepackage{epstopdf}
\usepackage{epsfig}
\usepackage{color}
\usepackage{lscape}
\DeclareGraphicsExtensions{.jpg,.pdf,.png,.eps,.ps}
\graphicspath{{FIGURES/}}

\newcommand{\ltsima}{$\; \buildrel < \over \sim \;$}
\newcommand{\ltsim}{\lower.5ex\hbox{\ltsima}}
\newcommand{\um}{$\mu \mathrm{m}$}
\newcommand{\beq}{\begin{equation}}
\newcommand{\eeq}{\end{equation}}
\newcommand{\smax}{\ensuremath{S_\mathrm{max}}}

\newcommand{\bfc}{\ensuremath{{\mathbf C}}}
\newcommand{\bfr}{\ensuremath{{\mathbf r}}}

\begin{document}

\submitted{Published in The Astrophysical Journal}

\title{A method for individual source brightness 
estimation in single- and multi-band data.}

\shorttitle{A method for individual source brightness estimation}


\author{
T. M. Crawford,\altaffilmark{1,2}
E. R. Switzer,\altaffilmark{1}
W. L. Holzapfel,\altaffilmark{3}
C. L. Reichardt,\altaffilmark{3}
D. P. Marrone,\altaffilmark{1,2,4} and
J. D. Vieira\altaffilmark{1,5,6}
}

\altaffiltext{1}{Kavli Institute for Cosmological Physics,
University of Chicago,
5640 South Ellis Avenue, Chicago, IL 60637}
\altaffiltext{2}{Department of Astronomy and Astrophysics,
University of Chicago,
5640 South Ellis Avenue, Chicago, IL 60637}
\altaffiltext{3}{Department of Physics,
University of California,
Berkeley, CA 94720}
\altaffiltext{4}{Jansky Fellow, National Radio Astronomy Observatory}
\altaffiltext{5}{Department of Physics,
University of Chicago,
5640 South Ellis Avenue, Chicago, IL 60637}
\altaffiltext{6}{Enrico Fermi Institute,
University of Chicago,
5640 South Ellis Avenue, Chicago, IL 60637}

\shortauthors{Crawford, et al.}

\email{tcrawfor@kicp.uchicago.edu}

\begin{abstract}
We present a method of reliably extracting the flux of individual
sources from sky maps 
in the presence of noise and a source population in which number counts
are a steeply falling function of flux.  The 
method is an extension of a standard Bayesian 
procedure in the millimeter/submillimeter literature.
As in the standard method, the 
prior applied to source flux measurements is derived from 
an estimate of the source counts as a function of flux, $dN/dS$.
The key feature of the new method 
is that it enables reliable extraction of properties of 
{\it individual} sources, which previous methods in the literature do not.
We first present the method for extracting
individual source fluxes from data in a single observing band, 
then we extend the method to multiple bands, including 
prior information about the spectral behavior of the source population(s).  
The multi-band estimation technique is particularly relevant for 
classifying individual sources into populations according to their spectral behavior.
We find that proper treatment of the correlated prior information between 
observing bands is key to avoiding significant biases in estimations of 
multi-band fluxes and spectral behavior, biases which lead to 
significant numbers of misclassified sources.
We test the single- and multi-band versions of the method 
using simulated observations with observing parameters
similar to that of the South Pole Telescope data used in \citet{vieira09}.
\end{abstract}

\section{Introduction}
\label{sec:intro}

The bias incurred when using noisy measurements to estimate 
source counts as a function of flux 
has been a recognized issue since at 
least the time of \citet{eddington13}.  The seminal work 
of \citet{scheuer57} on the topic was crucial to reconciling 
apparently conflicting radio source count measurements 
and establishing the model of an evolving universe
\citep[e.g.,][]{longair66}.

Accounting for this bias becomes particularly important when the 
source population under investigation has counts that are a steeply 
falling function of flux, as is the case for sources in the $1$-to-$100$~mJy 
range at millimeter/submillimeter (mm/submm) wavelengths.
Sources in this flux and wavelength range have been of particular recent
interest to astronomers, astrophysicists, and cosmologists, primarily due to the 
recently discovered population of high-redshift starburst galaxies 
(see \citet{blain02} for a review), which has been shown
to make up a substantial fraction of the cosmic infrared background 
\citep{lagache05} and which provides strong tests for theories of galaxy and 
star formation.  Both Bayesian 
and frequentist methods have been developed for
dealing with this bias in mm/submm surveys \citep[e.g.,][]{coppin05,maloney05}.  
However, as will be discussed in 
Sec.~\ref{sec:oneband}, the methods developed so far are not 
appropriate for estimating properties of individual sources.  

Dealing with this bias is further complicated when source properties are 
estimated from data in multiple wavelength bands.
\citet{mason09}, for example, used a maximum-likelihood technique 
to make an unbiased 
estimate of the spectral index distribution of a
population of sources from two-band centimeter-wave data, but they did not 
attempt to estimate spectral properties of individual sources.
When multi-band data have been used to estimate properties 
of individual sources, the data in different bands have generally 
been treated as independent, despite the fact that the information 
used to correct the biased flux measurements is highly correlated 
between bands~\citep[e.g.,][]{greve08}.  As will be shown in 
Sec.~\ref{sec:multiband}, ignoring these correlations can result 
in severely misestimated source properties.

In this work, we propose a reliable, minimally biased estimator for the 
single-band brightness of individual sources.  We then
extend this formalism to the simultaneous estimation of individual 
source properties over many bands, parameterized as either
the source brightness in each band or as the brightness in one 
band and the spectral behavior between bands.  The multi-band
implementation of the method explicitly accounts for
correlations in the information used to correct for flux bias in 
the individual bands.
This work was motivated by the extragalactic sources detected 
in multi-band South Pole Telescope (SPT) data, which are 
described in \citet[][hereafter V09]{vieira09}; however, we believe 
the method is directly applicable to other current or planned 
mm/submm experiments.  We also believe the method should 
be appropriate for use in single- and multi-band surveys
at other wavelengths, with the caveat that source variability could compromise 
the multi-band version of the method if the observations in different 
bands are not simultaneous.

\section{Single-band flux estimation}
\label{sec:oneband}

The estimation of source brightness in mm/submm surveys
is complicated by the fact that, at these wavelengths,
the number density of sources as a 
function of source flux is expected to be very steep.  
As a result, 
the measured flux of a source detected 
in a mm or submm survey will almost certainly suffer a positive
bias, often referred to as ``flux boosting."
In this work, we define flux boosting as the increased
probability that a source we measure
to have flux $S$ is really a dimmer source plus a positive noise
fluctuation over the probability that it is a brighter
source plus a negative noise fluctuation.  This asymmetric
probability distribution means that naive estimates of source
fluxes will be biased high.\footnote{This phenomenon is closely
related to what is referred to in the literature as ``Eddington bias"
\citep[e.g.,][]{teerikorpi04}; however, the consensus use of that term in 
the literature is to describe the bias introduced to estimation of source counts
vs.~brightness, not on the estimated brightness of individual sources.  This
usage is consistent with the original work of \citet{eddington13}, and the 
distinction we draw was also pointed out in (among others) \citet{hogg98}
and \citet{coppin06}.}   In the literature of mm/submm source surveys, 
one standard method for 
dealing with this problem \citep[e.g.,][]{coppin05} is to construct a posterior
probability distribution for the intrinsic flux of each detection, as
suggested by (among others) \citet{jauncey68} and \citet{hogg98}.  
According to standard Bayesian reasoning, this posterior distribution is given by
\beq
\label{eqn:bayes1}
P(S_i|S_m) \propto P(S_m|S_i) P(S_i),
\eeq
where $S_i$ is the intrinsic, true flux of the detected object, and
$S_m$ is the measured flux.  $P(S_m|S_i)$ is the likelihood of
measuring a flux $S_m$ given a true flux $S_i$, which in the simplest
case is a Gaussian distribution centered on $S_i$ with width
$\sigma_n$, determined solely by the noise in the maps from which
sources are being extracted.  $P(S_i)$ is the prior probability of a
source with intrinsic flux $S_i$, which is proportional to the
differential number counts vs.~flux, $dN/dS$.  

However, the formulation in Eqn.~\ref{eqn:bayes1} implicitly assumes 
that each detection corresponds to (at most) one real source.  This is
equivalent to assuming that the instrument used in the survey has
infinite resolution.  In reality, there will always be some
possibility that a resolution element containing a 
source above the detection threshold 
will also contain one or more fainter 
sources which will contribute to the measured flux in that resolution
element.  As a consequence, even in a noiseless measurement, the 
probability of measuring a particular value of flux within a finite resolution 
element is not identical (in general) to the probability that a single source 
of that flux exists within the resolution element.  
In much of the mm/submm literature (e.g., \citealt{coppin05} 
and \citealt{austermann09}), $P(S_p)$ (the probability of the total 
astrophysical flux $S_p$ within a pixel) and $P(S)$ (the
probability of finding a source of flux $S$ within the solid angle of
a pixel) are used interchangeably.  
In other words, what is being reported when this method is used
is not the distribution of intrinsic source fluxes (which is a property 
solely of the source population) but rather a distribution of pixel fluxes, which 
is also dependent on the survey instrument.  
When calculating source counts 
using this method, it is possible to use simulations to account for this instrumental 
dependence or demonstrate that it has a negligible effect 
\citep[e.g.,][]{austermann09}; 
however, the detailed shape of the posterior
flux distribtutions for individual sources can still be affected, particularly
at low significance.

One approach which avoids these particular complications is to find the 
underlying (intrinsic) number counts model that is consistent with the observed 
pixel flux distribution (as in, e.g., \citealt{maloney05} and 
\citealt{patanchon09}) or with the 
observed counts as a function of raw flux (as in 
``Reduction C" in \citealt{coppin06}).  However, these methods 
are incapable of estimating properties of individual sources.  
For this purpose, we extend the traditional Bayesian method to describe 
the properties of only one source in a given resolution element.  Instead of 
defining the posterior with respect to the (intrinsic) total flux in a resolution element 
or pixel, we define the posterior with respect to the intrinsic flux of the single 
brightest source in the resolution element.  The posterior probability 
distribution for this quantity is
\beq
\label{eqn:bayes}
P(\smax|S_{p,m}) \propto P(S_{p,m}|\smax) P(\smax),
\eeq
where $P(\smax|S_{p,m})$ is the posterior probability that the true flux of the
brightest source in a pixel is \smax \ given that we have measured the
total flux in that pixel to be $S_{p,m}$; $P(S_{p,m}|\smax)$ is the
likelihood of measuring flux
$S_{p,m}$ in a pixel given that the brightest source in that pixel has
flux \smax; and $P(\smax)$ is the prior probability that the brightest
source in that pixel has flux \smax. 

\subsection{Prior Probability $P(\smax)$}
\label{sec:psmax}
The prior probability $P(\smax)$ can be expressed as the probability
that one source of flux \smax \ exists in that pixel times the
probability that zero sources brighter than \smax \ exist in that pixel.  As
mentioned previously, the probability that within a
pixel there exists a source of flux $S$ is proportional to the
differential number counts (per unit flux per unit solid angle)
$dN/dS$.  Under the assumption of purely Poisson statistics, 
the probability that zero sources above \smax \ exist in a pixel is
\begin{eqnarray}
P(N>\smax = 0) &=& \frac{\mu^n}{n!}e^{-\mu}; \ n=0 \\
\nonumber &=& e^{-\mu},
\end{eqnarray}
where $\mu$ is the mean number of sources above \smax \ in pixels 
of size $\Delta \Omega_p$:
\beq
\mu(\smax) = \Delta \Omega_p \ \int_{\smax}^{\infty} \frac{dN}{dS} dS.
\eeq
In other words, $P(\smax)$ should look like $dN/dS$ with an
exponential suppression at low $S$ (where $\mu(S)$ will be large):
\beq
\label{eqn:psmax}
P(\smax) \propto \frac{dN}{dS} \bigg |_{S=\smax} 
\exp \left ( - \Delta \Omega_p \int_{\smax}^{\infty} \frac{dN}{dS^\prime} dS^\prime \right ).
\eeq

We note that this choice of prior is natural in the context of 
characterizing individual source properties and avoids certain 
complications associated with the standard choice of prior in 
the literature, which is the pixel probability distribution that 
would be obtained from analyzing noiseless
observations of a sky with the assumed underlying source 
distribution \citep[e.g.,][]{coppin05,austermann09}.  In particular, 
because point-source analyses of mm/sub-mm data almost 
inevitably involve spatially high-pass-filtering the data to remove
large-scale noise and astronomical signals, 
the standard prior can take on negative flux values.
This calls into question how such a prior can be 
describing the probability 
of the intrinsic flux of an astrophysical object.

\subsection{Likelihood $P(S_{p,m} | \smax)$}
\label{sec:condprob}
The total flux in a pixel given that the brightest source in that
pixel has flux \smax \ will be a sum of three contributions: the
source at \smax, a contribution from instrumental or atmospheric noise
in the survey, and a contribution from sources fainter than 
\smax.\footnote{In reality, there will also be contributions from
diffuse signals such as galactic dust or primary CMB anisotropy.  
However, these 
diffuse contributions will likely have been filtered out of maps 
used for source detection, so we neglect them here.}
We can
think of these each having its own probability distribution, and the
probability of the sum of their contributions to the flux in a pixel
will be the convolution of the individual distributions (by the
addition theorem for probabilities $P(u=x+y) = \int P_x(x) P_y(u-x)
dx$).  We will assume that the contribution 
from instrument noise and atmosphere
is well approximated by a Gaussian distribution:
\beq
P(S_{p,m},\text{noise-only}) = \frac{1}{\sqrt{2 \pi \sigma_n^2}} 
e^{-S_{p,m}^2/2\sigma_n^2},
\eeq
where $\sigma_n$ is the width of the combined instrument and atmosphere noise
distribution.  The contribution from sources fainter than \smax \ is 
given by the probability of total flux within a pixel, knowing that
there are no sources in that pixel brighter than \smax.
The probability for flux in a pixel given $dN/dS$ --- the so-called 
``deflexion probability'' or $P(D)$ ---
is worked out in \citet{scheuer57}, and, under the 
assumption of pure Poisson statistics is
\beq
P(S_{p,m},\text{noise-free}) = FT\{e^{\left[r(\omega) - r(0) \right]} \},
\eeq
where $FT\{\}$ denotes Fourier Transform, and $r(\omega)$ is the characteristic function of the probability of finding a source of flux $S$ in a pixel of solid angle $\Delta \Omega_p$:
\beq
\label{eqn:romega}
r(\omega) = FT \left \{ \frac{dN}{dS} \Delta \Omega_p  \right \}.
\eeq
To calculate the contribution of sources fainter than \smax \
under the constraint that there are no sources in a pixel greater than
\smax, we apply the above result using a $dN/dS$ distribution
that is truncated at \smax.  

To summarize, the flux in a pixel given that we have exactly one
source of flux \smax \ and zero sources above that flux will be be sum
of the contribution from the source at \smax, the contribution from
sources fainter than \smax, and the contribution from noise.  The probability 
distribution for this sum is the convolution of the individual distributions, so that
\begin{eqnarray}
\label{eqn:condprob}
&& P(S_{p,m} | \smax) = \\
\nonumber && \ \delta(\smax) \ast FT\{e^{\left[r(\omega) - r(0) \right]} \}
\ast \frac{1}{\sqrt{2 \pi \sigma_n^2}} e^{-S_{p,m}^2/2\sigma_n^2},
\end{eqnarray}
where ``$\ast$'' denotes convolution.  

\subsection{Beam and filtering effects}
The exact formulations in Eqns.~\ref{eqn:psmax} and \ref{eqn:condprob} 
only hold for 
a survey in which the only spatial filtering done involves binning into 
pixels.  For most real instruments, the situation is 
complicated by the instrument beam (or point-spread function) 
and by the time-domain and map-domain
filtering performed on the data.  The likelihood in 
Eqn.~\ref{eqn:condprob} needs to be modified to reflect the difference in how 
sources fainter than \smax \ contribute to a real instrumental beam compared to the top-hat
pixel considered earlier.  \citet{scheuer57} shows that Eqn.~\ref{eqn:romega} can 
be modified for finite-resolution experiments by defining:
\beq
\label{eqn:romega2}
r(\omega) = FT \left \{ \int \frac{dN}{dS_\mathrm{beam}} \frac{d \Omega}{B \left (\theta,\phi \right)}  \right \},
\eeq
where $B(\theta,\phi)$ is the angular response pattern of the instrument, and 
\beq
S_\mathrm{beam} = \frac{S}{B \left (\theta,\phi \right)}.
\eeq
This formulation can be extended to account for any filtering in the data analysis
by modifying the angular response function to include the filtering.

The prior in Eqn.~\ref{eqn:psmax} and the $\delta(\smax)$ term in Eqn.~\ref{eqn:condprob} 
must also be modified for a finite-resolution experiment.  We choose to define 
\smax \ for a finite-resolution experiment 
as the source that contributed the most flux in a resolution element.  That means that 
both equations must be replaced by integrals over the beam with \smax \ replaced 
by \smax \ times the beam response.  Also, for each value of the integrand in 
Eqn.~\ref{eqn:romega2}, the value at which $dN/dS$ is truncated in the $r(\omega)$ 
calculation will be different.  This quickly becomes computationally intractable, but 
here we are actually helped by the expected steepness of the source distribution.
If we define $S_{\mathrm{max},p}$ as the largest flux contribution in a resolution
element,  the probability that the source contributing that flux is a source of intrinsic
flux very close to $S_{\mathrm{max},p}$ that lies near the center of the beam is far larger 
than the probability that the contributing source is a much brighter source far off 
beam center.  This means we can approximate the correct versions of 
Eqs.~\ref{eqn:psmax} and \ref{eqn:condprob} (that include integrals over the beam) 
with the original, infinite-resolution versions, using a value for $\Delta \Omega_p$ 
that is roughly the area of the beam over which its response is near unity.
In comparing to the simulated observations described in Sec.~\ref{sec:simsoneband}, 
we use $1 \ \mathrm{arcmin}^2$, which is roughly the square of the full width at half maximum
of the beam used in the simulations.  

\subsection{Comparison with simulations}
\label{sec:simsoneband}
We use simulated observations of mock skies including point-source populations
drawn from model $dN/dS$ distributions and Gaussian-distributed noise 
to test the formalism developed in the previous sections.
Simulated observations were performed using three sets of observing 
parameters, each roughly consistent with the $2.0$~mm maps presented
in V09, which we use as a concrete example.  The three sets of observing 
parameters used are:  1) No spatial filtering beyond binning into $1$-arcmin pixels 
(and subtracting off the mean value of the map), 
noise consistent with the $2.0$~mm SPT map shown in  V09 
($\sim 1.4$~mJy rms); 2) Noise as in 1, 
but with spatial filtering similar to the real SPT $2.0$~mm maps in  V09;
3) Filtering as in 1, but with the noise level halved.
The source count model that was used to create the simulated maps 
is the sum of the \citet{negrello07} 850~\um \ counts for dust-dominated  
sources (scaled to $2.0$~mm as in  V09) 
and the \citet{dezotti05} counts for synchrotron-dominated sources.  

From these simulated observations, we extract the true, underlying 
posterior flux PDF, $P(\smax|S_{p,m})$, for three different values of $S_{p,m}$
in each observing configuration.
The true, underlying posterior flux PDF for a given value of $S_{p,m}$ was 
estimated by taking each source detected in
the simulated maps
with measured flux within $\delta S$ of $S_{p,m}$, finding every source 
in the true, underlying source population that was associated with that 
detection, and recording the flux of the brightest associated source as \smax \ for 
that detection.  In the pixel-only cases, sources were considered associated 
with a detection if they were in the same pixel as the detection; in the 
beam-and-filtering case, they were considered associated if they were 
within 1 arcmin of the position of the detection.
The true, underlying $P(\smax|S_{p,m})$ for that value of
$S_{p,m}$ is then simply the histogram of the \smax \ values assigned to
all the detections with measured flux within $\delta S$ of $S_{p,m}$.  A total 
of 200 simulated $100 \ \mathrm{deg}^2$ maps were used to construct 
the histogram.

\begin{figure*}[h]
 \begin{center}
\epsfig{file=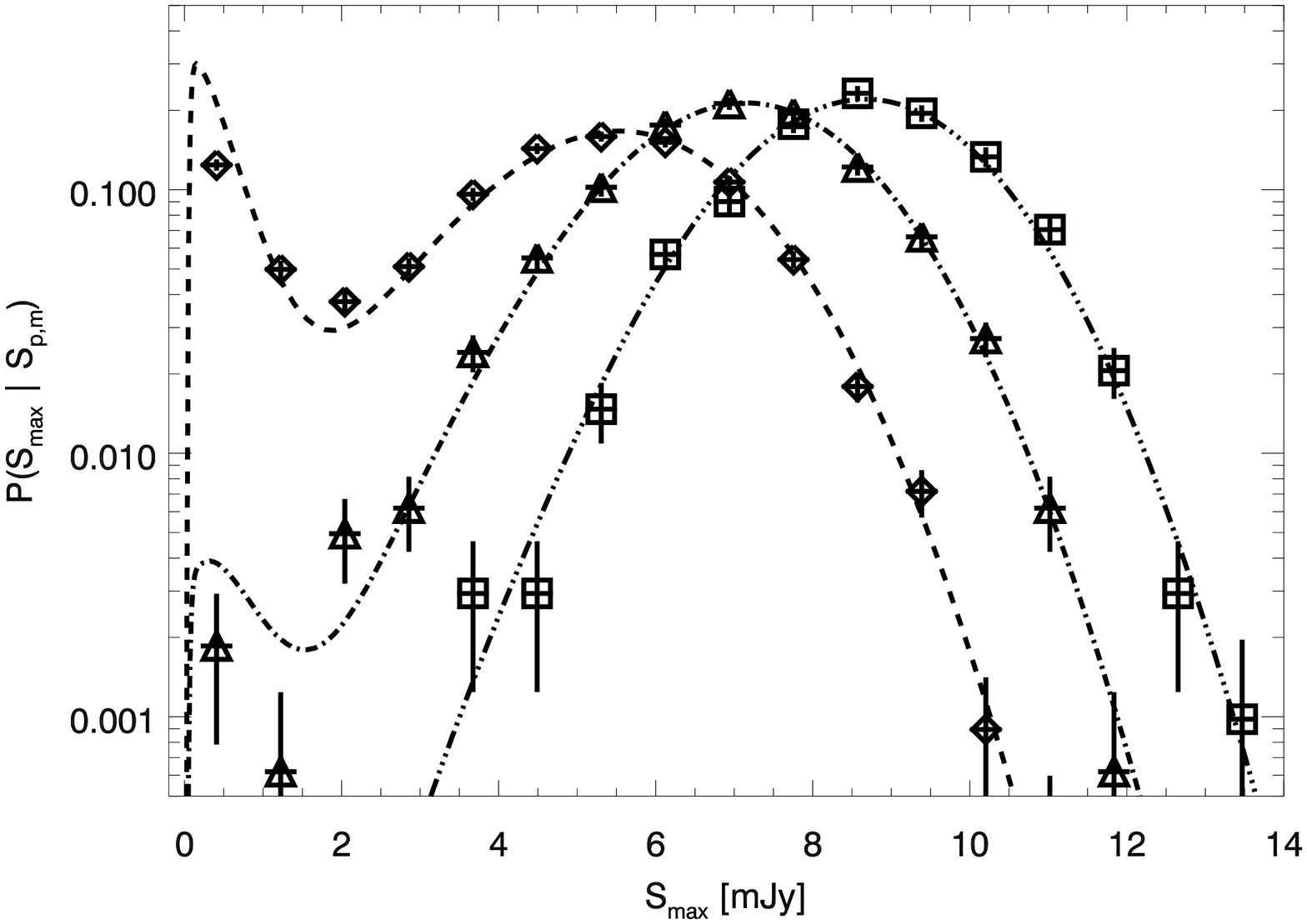, width=8cm} 
\epsfig{file=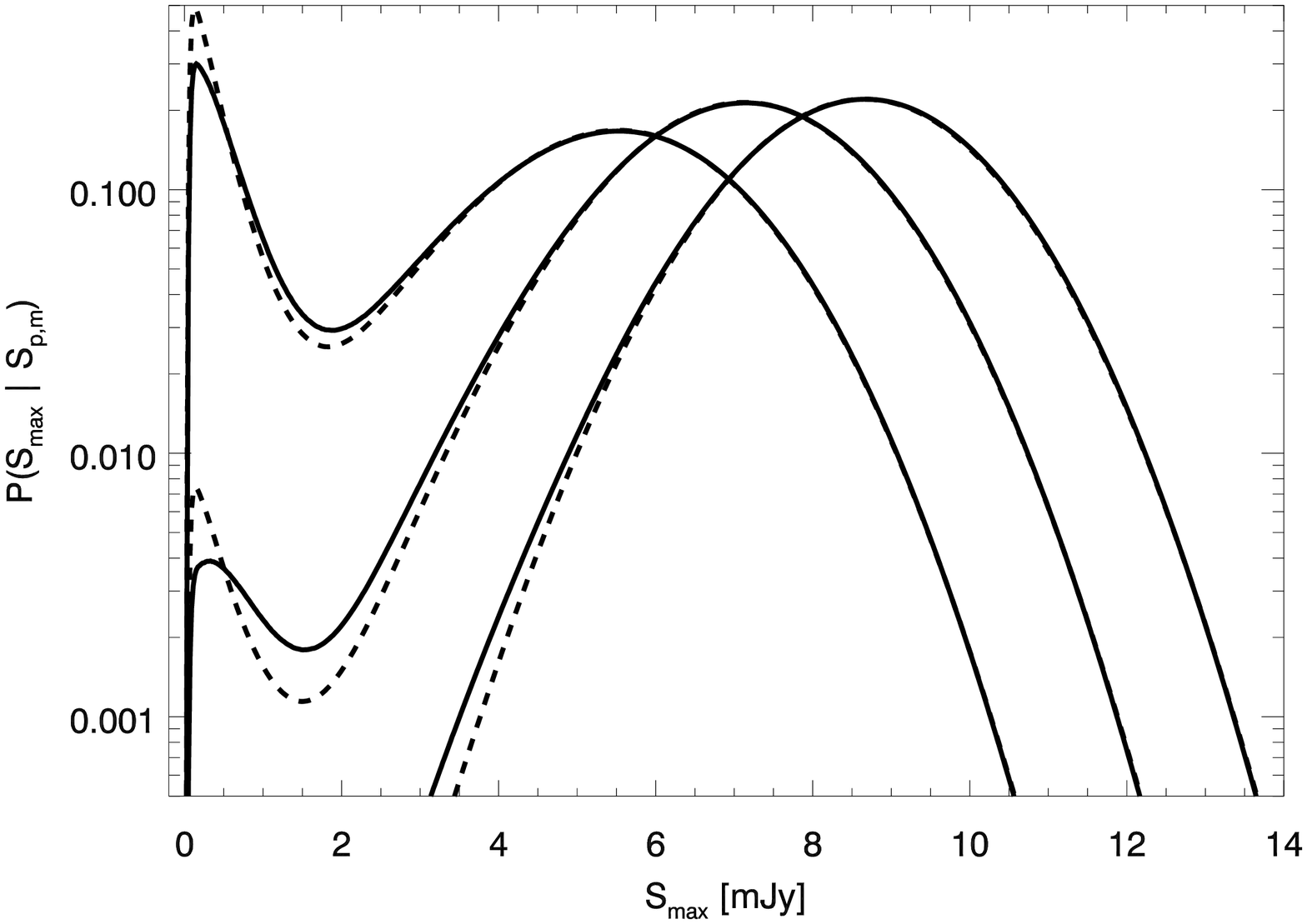, width=8cm} 
\epsfig{file=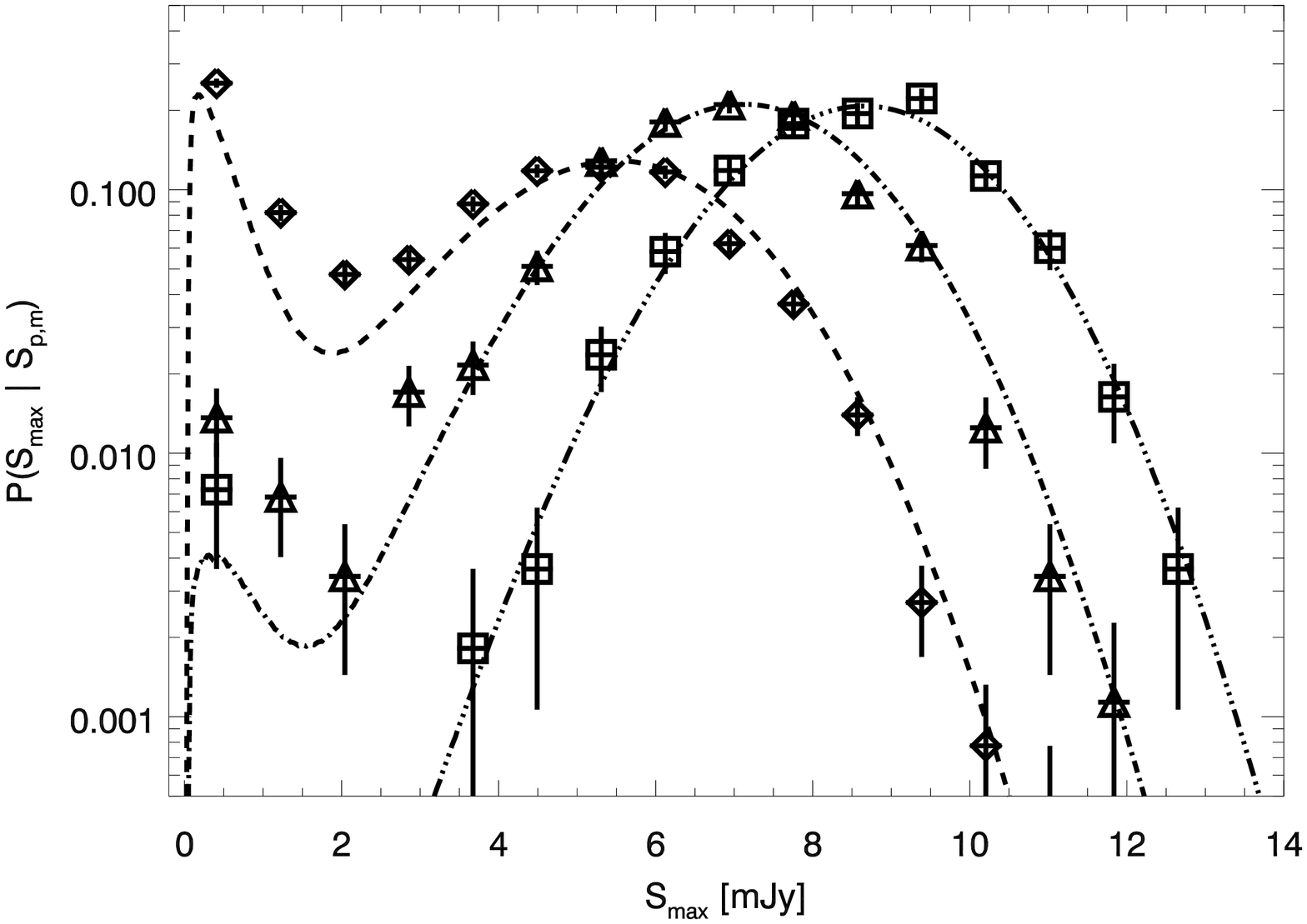, width=8cm} 
\epsfig{file=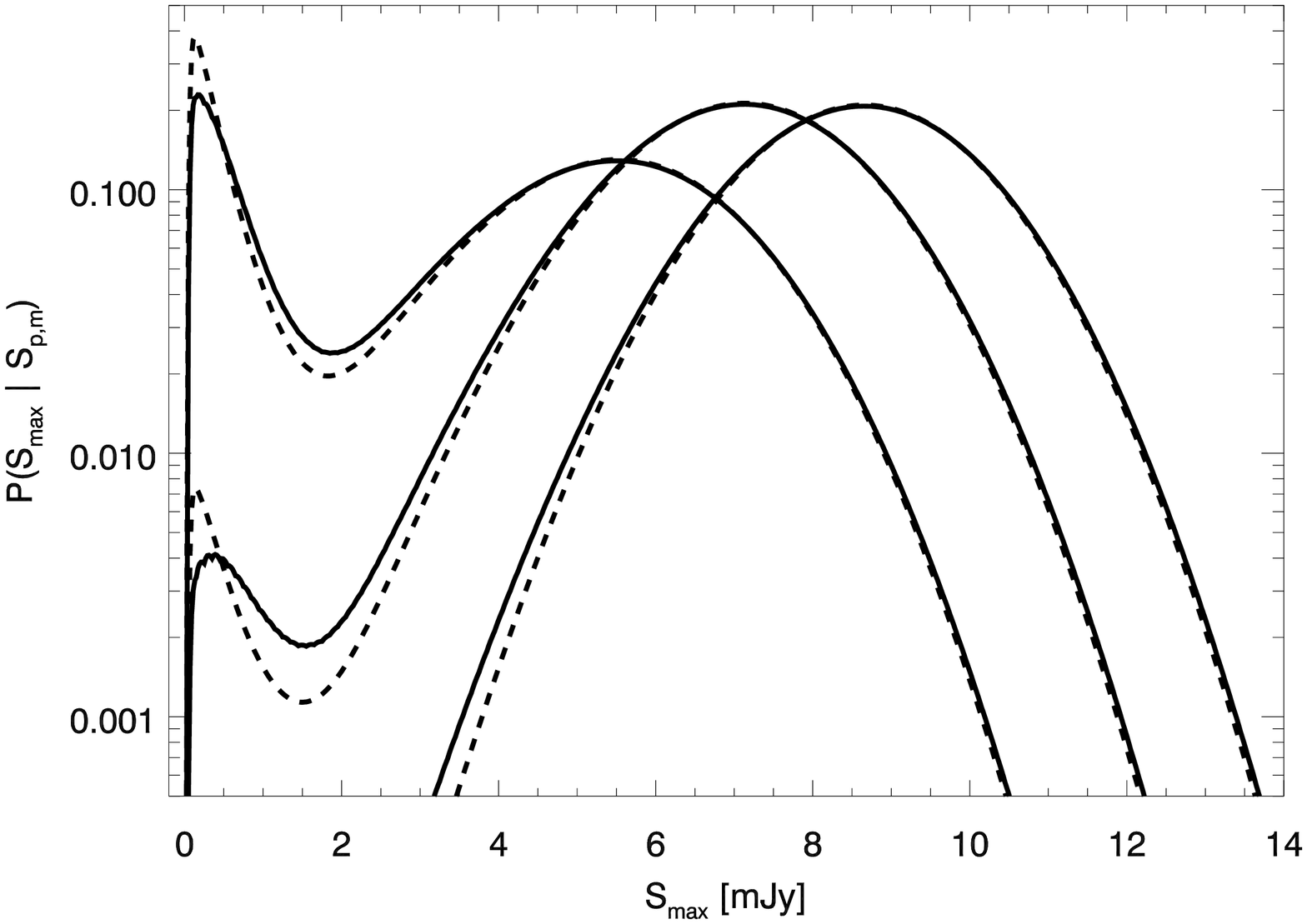, width=8cm} 
\epsfig{file=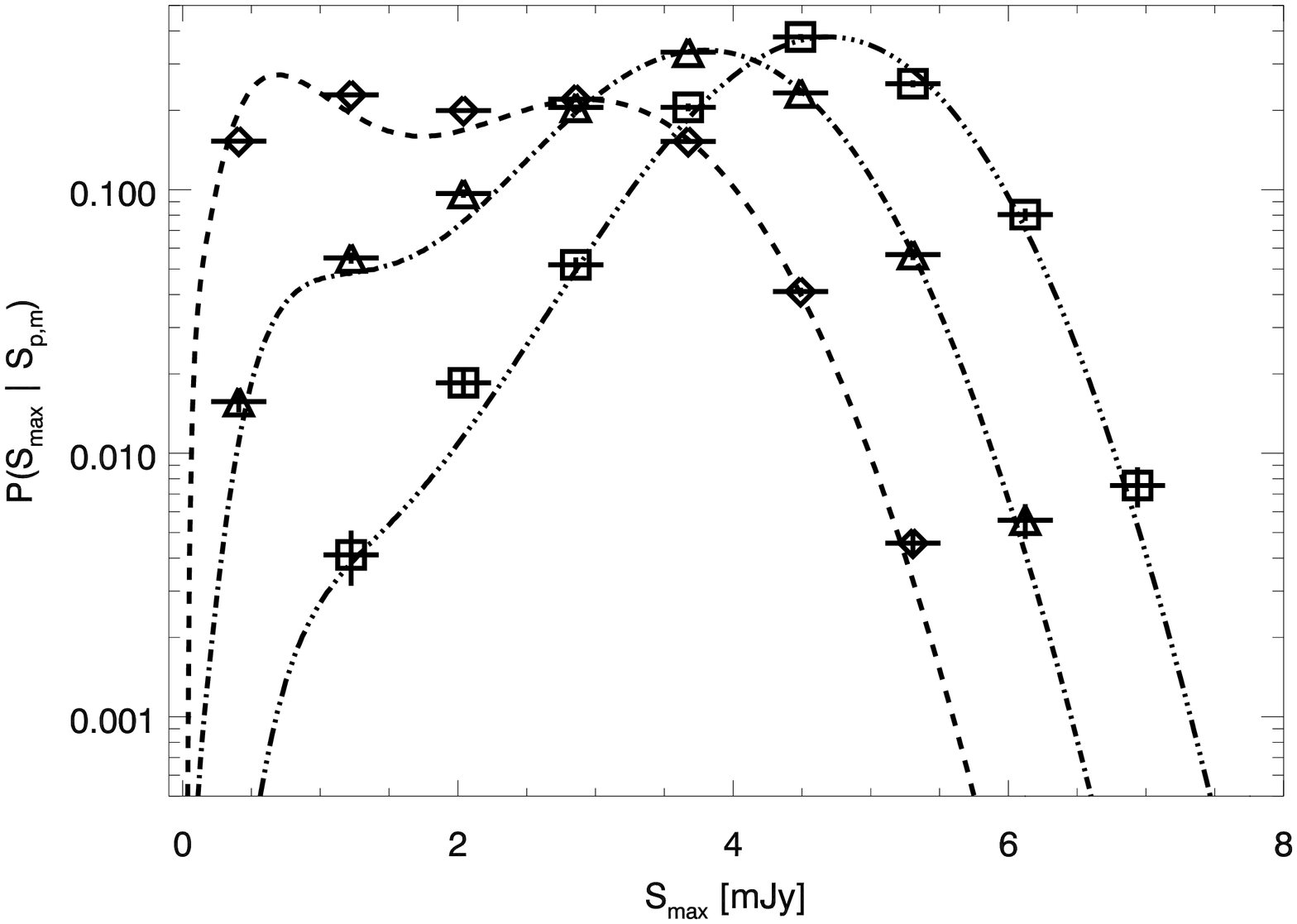, width=8cm} 
\epsfig{file=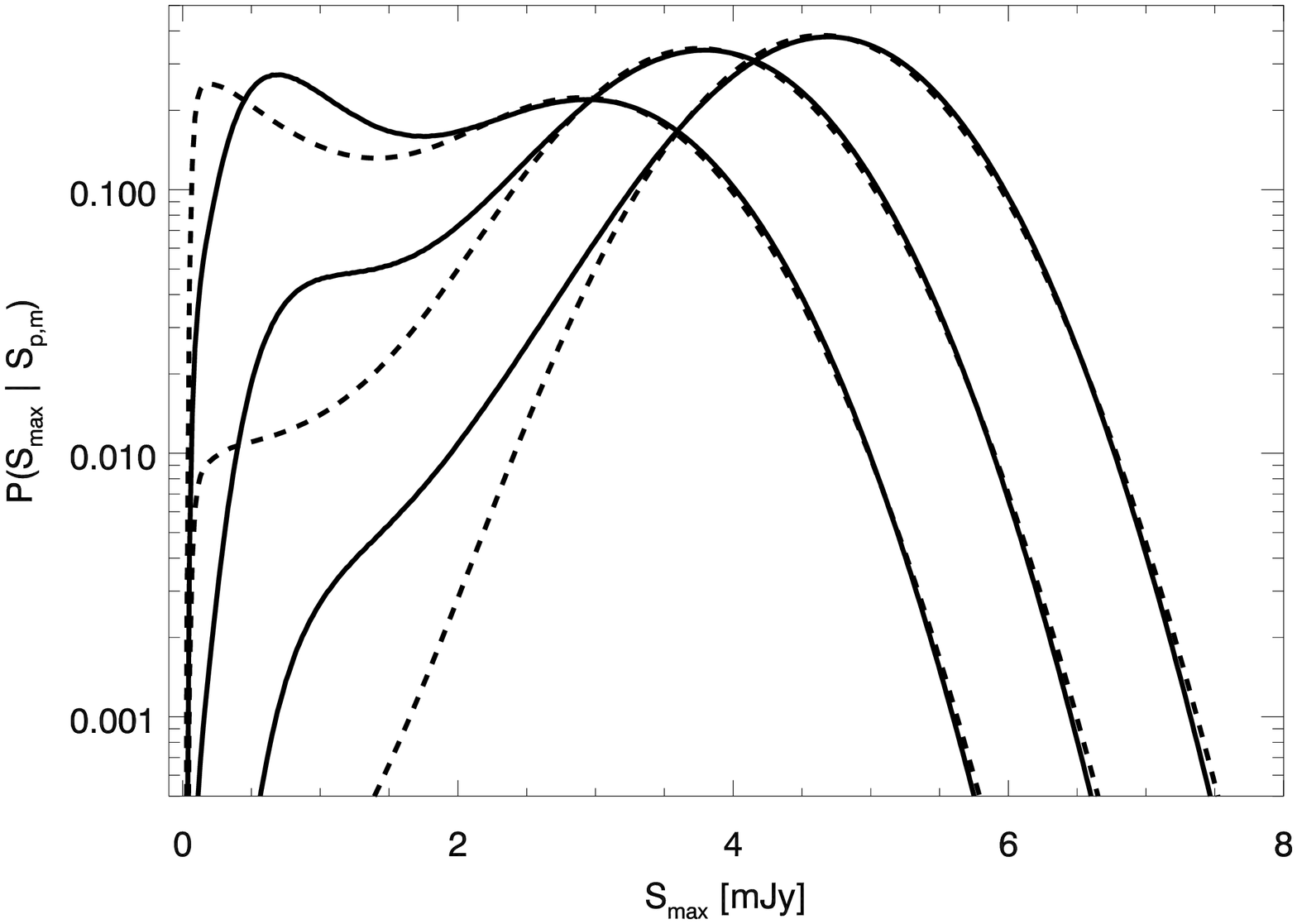, width=8cm} 
\end{center}
\caption{\scriptsize
{\bf Left Panels}: True, underlying posterior flux PDF $P(\smax|S_{p,m})$
(extracted from 
simulated observations and shown by symbols with error bars) and 
calculated values for that PDF (using the procedure
outlined in Sec.~\ref{sec:oneband} and shown by lines).
In the top two rows, the raw flux values for which the posterior PDFs are calculated 
are $6.3$~mJy ({\it diamond symbols and dashed line}), $7.7$~mJy ({\it triangle
symbols and dot-dashed line}), and $9.1$~mJy ({\it square symbols and 
triple dot-dashed line}).  In the bottom row, those raw flux values are 
$3.4$, $4.1$, and $4.9$~mJy.  In both cases, these raw flux values 
correspond to detection significance levels of $4.5 \sigma$, $5.5 \sigma$, 
and $6.5 \sigma$.
Vertical error bars on the extracted PDFs are from Poisson statistics; 
horizontal error bars show $\delta S=0.2$~mJy, the width of the flux bins used
to construct the $P(\smax|S_{p,m})$ histogram from the simulated observations
(see text for details).
{\bf Right Panels}: Gaussian likelihood approximation to the posterior PDF
calculation.  Solid lines are from the full calculation and are identical
to the lines in the left panel; dashed lines are calculated using the 
Gaussian approximation to the likelihood 
$P(S_{p,m} | \smax)$ shown in Eqn.~\ref{eqn:condprobgauss}.
{\bf Top Row}:  binning into $1$-arcmin pixels only, $1.4$~mJy noise rms;
{\bf Middle Row}:  full beam and filtering, $1.4$~mJy noise rms;
{\bf Bottom Row}:  binning into $1$-arcmin pixels only, $0.7$~mJy noise rms;
\label{fig:psmax_smeas}}
\end{figure*}

Fig.~\ref{fig:psmax_smeas} shows the posterior flux PDFs extracted from 
the simulated observations and the 
calculated values of those posterior flux PDFs (using the formalism 
developed in the previous sections). 
The source model used to calculate the prior probability $P(\smax)$ 
is the same model used to generate the mock point-source skies.
The values of $S_{p,m}$ for which posterior flux PDFs are shown  
correspond to detection significance values of $4.5 \sigma$, $5.5 \sigma$, and 
$6.5 \sigma$ in each set of simulated observations.  These detection significance values
correspond to raw flux values of $6.3$, $7.7$, and $9.1$~mJy in the full-noise simulations 
and $3.4$, $4.1$, and $4.9$~mJy in the halved-noise simulations.
(The detection significance in the halved-noise simulations for a given raw
flux value are not exactly twice those in the full-noise simulations because 
of the contribution of background sources to the map rms.)

For comparison with the output of the pixel-only simulated observations, 
the posterior flux PDF was calculated exactly using the equations in 
Sec.~\ref{sec:psmax} and Sec.~\ref{sec:condprob}.  The top left and bottom 
left panels of Fig.~\ref{fig:psmax_smeas} show that the calculated
posterior PDF values for these cases
are consistent with the simulation output to within the assumed Poisson errors.
In the full-beam-and-filtering case, 
$r(\omega)$ was calculated exactly, but the approximation described 
in the previous section was used for the prior and for the $\delta(\smax)$ term in 
Eqn.~\ref{eqn:condprob}.  The small but statistically measurable discrepancies 
between the simulated and calculated PDFs in the full-beams-and-filtering case
(middle left panel of Fig.~\ref{fig:psmax_smeas}) are due to the imperfect nature of this 
approximation.  

\subsection{Gaussian likelihood approximation}
\label{sec:gauss1}

The most computationally intensive step in estimating the posterior 
flux probability distribution $P(\smax | S_{p,m})$ is calculating the 
contribution from the other sources in the resolution element.  For 
deep, single-dish mm/submm surveys, in which the angular 
resolution varies from $\sim 10$ arcseconds to a couple of arcminutes, this 
contribution will be dominated by the background of sources
below the confusion limit (the regime in which there are many 
sources per resolution element).  
By the central limit theorem, as the number of sources 
per resolution element becomes large, the distribution of total flux from these 
sources will approach a Gaussian.  If we use 
this fact to approximate the background source contribution to the 
posterior as another Gaussian-distributed noise source, the calculation
of the likelihood in Eqn.~\ref{eqn:condprob} becomes 
trivial:
\beq
P(S_{p,m} | \smax) = \frac{1}{\sqrt{2 \pi \sigma_\mathrm{tot}^2}} 
e^{- \left ( S_{p,m} - \smax - \overline{S_p} \right)^2/2\sigma_\mathrm{tot}^2},
\label{eqn:condprobgauss}
\eeq
where $\overline{S_p}$ is the mean contribution from sources in the map, 
and $\sigma_\mathrm{tot}$ is the quadrature 
sum of the instrumental and atmospheric noise and the fluctuations in the 
source background:
\beq
\sigma^2_\mathrm{tot} = \sigma_n^2 + \sigma_\mathrm{sources}^2.
\eeq

There will always be a high-flux tail to the $P(D)$ distribution, 
because some resolution elements in the survey will have more than 
one source brighter than the confusion limit.  However, for steep source 
populations, the mean number of sources contributing to the total 
flux in a pixel will be large, and the tails will be less important.  
Furthermore, because the background source contribution adds 
in quadrature to the instrument and atmospheric noise contribution
(which is likely to be very well approximated by a Gaussian),  
non-Gaussianity in the source contribution will manifest itself in the 
total $P(D)$ distribution only if it is the dominant source of noise
in a survey.  For a survey like the SPT, this is not the case, and the 
Gaussian likelihood approximation holds very well, as shown in the 
top and middle right panels of Fig.~\ref{fig:psmax_smeas}.  However, a survey
that was a factor of two deeper at the same wavelength and resolution would incur 
much more significant errors by adopting the Gaussian likelihood 
approximation, as shown in the bottom right panel of 
Fig.~\ref{fig:psmax_smeas}.

It is worthwhile to note that adopting the Gaussian likelihood 
approximation removes one of the differences between the flux 
estimation method presented here and the standard method in the 
literature.  In the Gaussian likelihood approximation, the implicit 
assumption is made that the only contributions to the 
measured flux in a pixel are instrumental and atmospheric noise, a
single bright source, and a source background that acts as another 
symmetric noise source.  In this case, the only difference between the likelihood
of measuring some flux in a pixel given the flux of the brightest source 
in that pixel and the likelihood of measuring some flux in a pixel given the intrinsic 
total flux in that pixel is the mean contribution from the source background.  
Since all of the mm/submm experiments under consideration here 
are differential (i.e., not sensitive to the mean brightness on the sky), the
two likelihoods are effectively identical.  However, a significant difference
remains between the priors used in the two methods, as discussed in 
Sec.~\ref{sec:psmax}.

\subsection{Estimating source counts with a source-count prior}

If the individual source fluxes estimated with this method (or with the
standard method in the literature) are 
used in estimating source counts (as they are in 
V09 and most submm analyses), 
one might object that a prior probability has 
been assumed for the very quantity that one is attempting to 
measure.  It is important to remember that the source-count 
prior is applied probabilistically to determine each source's flux, 
so the estimated source counts will only resemble the prior if
the measurements have no constraining power.  From the posterior
flux PDFs shown in Fig.~\ref{fig:psmax_smeas}, it is clear that 
for the combination of data and prior used in these simulated 
observations, the data are providing the bulk of the information
in the flux measurement down to the lowest detection significance
shown ($4.5 \sigma$).  For a real survey, plots of individual posterior 
flux PDFs such as these are useful for determining where the 
reported source counts contain significant new information and 
where they simply reproduce the prior.  Another useful check on
the constraining power of the data is to vary the source count prior
and confirm that the estimated counts do not change significantly 
(as in \citealt{scott08} and V09).


\section{Multi-band source flux estimation}
\label{sec:multiband}

The posterior flux estimation method described in Sec.~\ref{sec:oneband} 
implicitly assumes that both the data and the prior knowledge of the source
distribution are restricted to a single observing band.  The situation will 
inevitably be more complicated when both data and priors are available 
in multiple bands.
In full generality, the task at hand is now to simultaneously 
estimate the posterior probability of the intrinsic source flux in multiple bands given
the measured flux in those bands and any prior information.  Using the 
formalism from Sec.~\ref{sec:oneband}, we can write the two-band case as
\begin{eqnarray}
\label{eqn:bayes2}
&&P(S_{\mathrm{max},1},S_{\mathrm{max},2}|S_{p,m,1},S_{p,m,2}) 
\propto  \\
\nonumber && P(S_{p,m,1},S_{p,m,2}|S_{\mathrm{max},1},S_{\mathrm{max},2})
P(S_{\mathrm{max},1},S_{\mathrm{max},2}).
\end{eqnarray}

In the mathematically simplest case, both the conditional and prior probabilities
on the right-hand side of Eqn.~\ref{eqn:bayes2} are uncorrelated between bands, 
so that the posterior probability distribution for flux in the two bands is simply the product of 
the two single-band distributions:
\begin{eqnarray}
\label{eqn:uncorr}
&&P(S_{\mathrm{max},1},S_{\mathrm{max},2}|S_{p,m,1},S_{p,m,2}) 
\propto  \\
\nonumber && 
P(S_{p,m,1}|S_{\mathrm{max},1}) P(S_{\mathrm{max},1})
P(S_{p,m,2}|S_{\mathrm{max},2}) P(S_{\mathrm{max},2}).
\end{eqnarray}
However, the assumption of uncorrelated prior information in the two bands 
will very rarely be valid.  The prior in each band will be derived from previous 
estimates of source counts at wavelengths and flux levels as near as possible 
to that band, using assumptions about spectral behavior to extrapolate to the 
bands of interest.  For bands that are reasonably close to each other in 
wavelength, the existing data that
go into estimating the priors in the two bands will almost certainly have some 
overlap.  This is not simply a matter of imperfect priors; in the limit of perfect 
prior knowledge of the source counts in both bands, the priors in the 
two bands would only be independent if the source populations measured 
in the two bands had zero overlap.  In the example of the SPT 
data at 1.4 and 2.0~mm analyzed in V09, the prior probability $P(\smax)$ in both bands 
is dominated in the few-mJy flux region by the dusty starburst population, 
with some contribution from synchrotron-dominated AGN.  It is clearly a 
poor approximation to assume that the priors in these two bands are uncorrelated.  

It is 
possible to construct the full two-dimensional prior probability 
$P(S_{\mathrm{max},1},S_{\mathrm{max},2})$; however, it will often be
more convenient to change variables and cast this prior probability 
in terms of the flux in one band and the expected spectral behavior 
of the sources contributing to the prior.  We define the spectral index 
$\alpha$ through the relation
\beq
S(\lambda_2) = S(\lambda_1) \ \left ( \frac{\lambda_2}{\lambda_1} \right )^{-\alpha}
\label{eqn:alpha}
\eeq
and write the two-dimensional prior as $P(S_{\mathrm{max},1},\alpha)$.
The only requirement for this prior to factor into independent priors 
$P(S_{\mathrm{max},1})$ and $P(\alpha)$
is that the spectral index distribution between the two bands not depend on flux.
Of course, in reality $P(\alpha)$ will depend somewhat on flux --- for example, in 
the SPT case, the brightest sources are mostly AGN, but near the confusion limit 
the source population is dominated by dusty galaxies.
In cases such as the analysis of SPT data in V09, in which estimating the spectral index 
distribution of the sources is one of the key goals of the analysis, the prior used 
will be sufficiently weak (V09 use a flat prior from $-3 \le \alpha \le 5$)
that the slight dependence on flux of the real $P(\alpha)$ is of no consequence.
Alternatively, we can accept the added
complexity of using the full two-dimensional prior on \smax \ and $\alpha$ --- assuming
sufficient data exist to meaningfully construct such a distribution.

No matter how we choose to construct the spectral index prior, we now have
the two-dimensional posterior PDF of flux in one band and 
$\alpha$:
\begin{eqnarray}
\label{eqn:psmax_alpha}
&&P(S_{\mathrm{max},1},\alpha|S_{p,m,1},S_{p,m,2}) 
\propto  \\
\nonumber && 
P(S_{p,m,1},S_{p,m,2}|S_{\mathrm{max},1},\alpha)
P(S_{\mathrm{max},1},\alpha).
\end{eqnarray}
If we choose, we can then transform this into a two-dimensional
posterior probability distribution for the flux in both bands:
\begin{eqnarray}
\label{eqn:psmax_smax}
&&P(S_{\mathrm{max},1},S_{\mathrm{max},2}|S_{p,m,1},S_{p,m,2}) = \\
\nonumber && 
P(S_{\mathrm{max},1},\alpha|S_{p,m,1},S_{p,m,2}) \ \frac{d \alpha}{d S_{\mathrm{max},2}},
\end{eqnarray}
where  $d \alpha / d S_{\mathrm{max},2}$ is derived from Eqn.~\ref{eqn:alpha}.

\subsection{Choice of detection band}
\label{sec:priorswap}
We note that as the prior on $\alpha$ is relaxed to a flat prior
between $-\infty$ and $\infty$ (independent of source flux), 
the posterior PDF for $S_{\mathrm{max},1}$ in
Eqn.~\ref{eqn:psmax_alpha} reduces to the single-band posterior of 
Eqn.~\ref{eqn:bayes}.  Meanwhile, the posterior 
PDF for $S_{\mathrm{max},2}$ becomes equal to the likelihood
$P(S_{p,m,2}|S_{\mathrm{max},2})$ --- i.e., there is effectively no prior on 
$S_{\mathrm{max},2}$.  This points out an apparent asymmetry between 
bands in our approach, namely that the two-band posterior flux PDF 
will depend on which band you choose to use prior source count 
information from --- call this the ``detection band" --- 
and which band you apply prior information to only 
through the combination of the prior on $\alpha$ and source-count prior
on the first band.  

The real issue is that $dN/dS$ in each band and a distribution in 
$\alpha$ are not three independent pieces of information; 
rather, the choice of $dN/dS$ in one band and a distribution in $\alpha$ 
uniquely specifies $dN/dS$ in the other band.  If the assumptions 
regarding $dN/dS$ in both bands and the distribution of $\alpha$ are 
internally consistent, then the two-band posterior flux PDF will be 
identical regardless of which band is chosen as the detection band.
However, if one of the goals of the analysis is to measure the distribution
of spectral indices, such a prior on $\alpha$ may be too restrictive.
In  V09, the choice was made to adopt a non-restrictive prior 
($-3 \le \alpha \le 5$) and analyze the data using each band in turn
as the detection band.  The differences in the final derived source 
counts between the two analyses were small compared to the 
statistical uncertainties.

\subsection{Variable sources}
\label{sec:variability}
The members of at least one of the populations expected to contribute 
to mm/submm source counts --- namely AGN --- are expected to have 
highly time-variable flux.  This complicates any estimate of 
source properties from data taken at different epochs.  For instruments 
that observe simultaneously in multiple wavelength bands --- as the SPT 
does --- this is not an issue, and we have ignored any effects of variability
on the method presented here.  This would be an issue, however, in 
extending this method to radio survey data.

\subsection{Gaussian likelihood approximation in the multi-band case}
As in the single-band case, the calculation of the likelihood 
$P(S_{p,m,1},S_{p,m,2}|S_{\mathrm{max},1},\alpha)$ is greatly simplified 
by the assumption that the instrumental and atmospheric noise and 
the contribution from sources below \smax \ are 
Gaussian-distributed.  While the full, non-Gaussian two-dimensional 
likelihood is in principle calculable --- by extending the 
result of \citet{scheuer57} into a multivariate Fourier operation --- that 
derivation is beyond the scope of this work.  It is also possible to 
estimate the full, non-Gaussian two-dimensional distribution by simulated 
observations.  For now, we will adopt the Gaussian likelihood approximation 
and use simulated observations to check its validity, as we did for the 
single-band results.

Under the Gaussian likelihood approximation, the two-band likelihood
(analogous to the single-band likelihood in Eqn.~\ref{eqn:condprobgauss})
is given by
\beq
P(S_{p,m,1},S_{p,m,2}|S_{\mathrm{max},1},\alpha) = \frac{\exp{\left( -\frac{1}{2} \ \bfr^T \bfc^{-1} \bfr\right)}}{2 \pi \sqrt{\det{\bfc}}},
\eeq
where \bfc \ is the noise covariance between the bands (including contributions 
from instrument noise, atmosphere, and sources fainter than \smax), and \bfr \ is the residual 
vector
\begin{eqnarray}
\label{eqn:residvec}
\bfr &=& \Bigl \{ S_{p,m,1} - S_{\mathrm{max},1} - \overline{S_{p,1}}, \\
\nonumber && \ S_{p,m,2} - S_{\mathrm{max},2} (\alpha) - \overline{S_{p,2}} \Bigr \} \\
\nonumber &=& \Bigl \{ S_{p,m,1} - S_{\mathrm{max},1} - \overline{S_{p,1}}, \\
\nonumber && \ S_{p,m,2} - S_{\mathrm{max},1} \left ( \frac{\lambda_2}{\lambda_1} \right )^{-\alpha} - \overline{S_{p,2}} \Bigr \}.
\end{eqnarray}

\subsection{Comparison to simulations}
\label{sec:simstwoband}
 
\begin{figure}[h]
\begin{center}
\epsfig{file=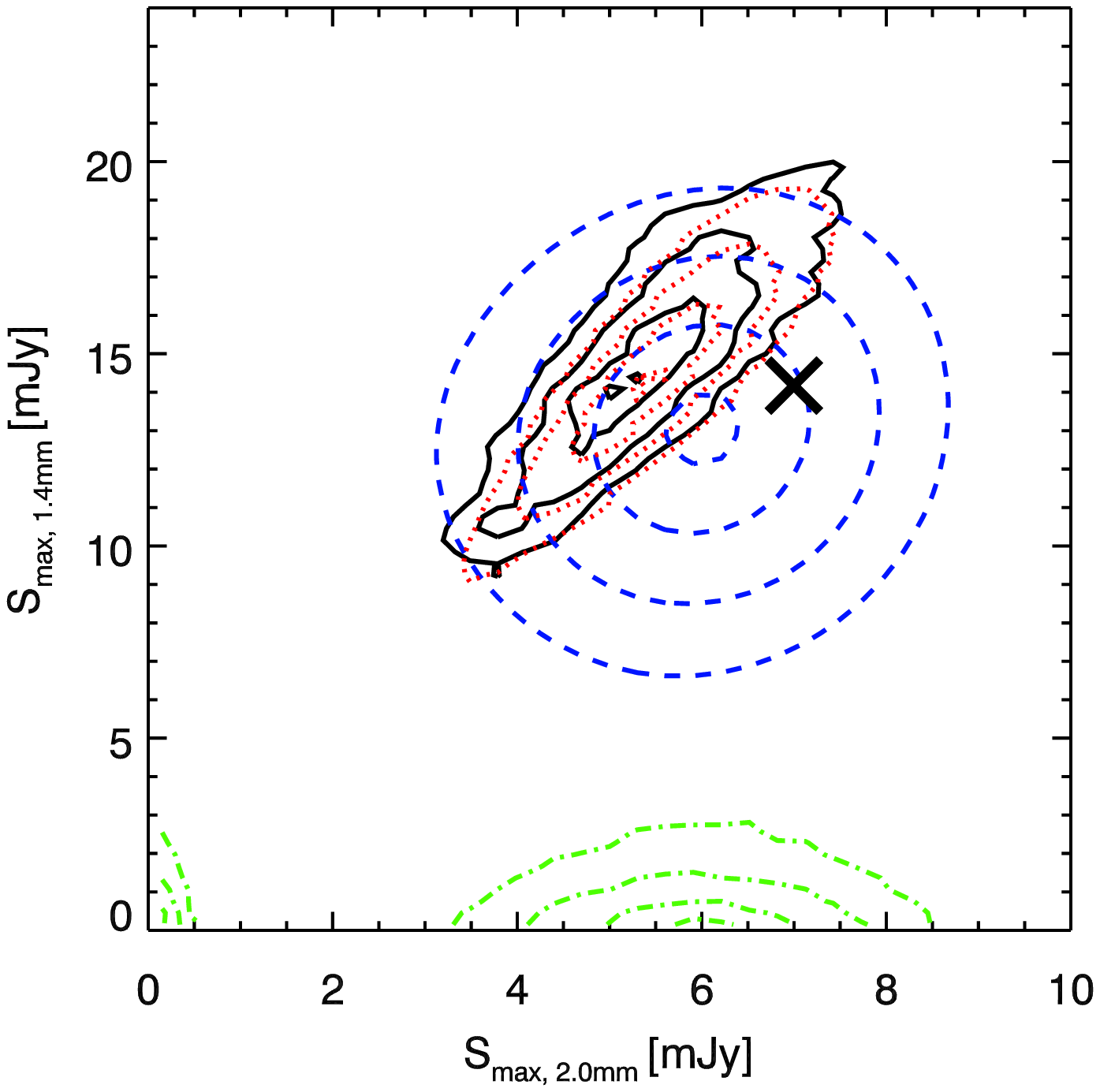, width=7cm} 
\epsfig{file=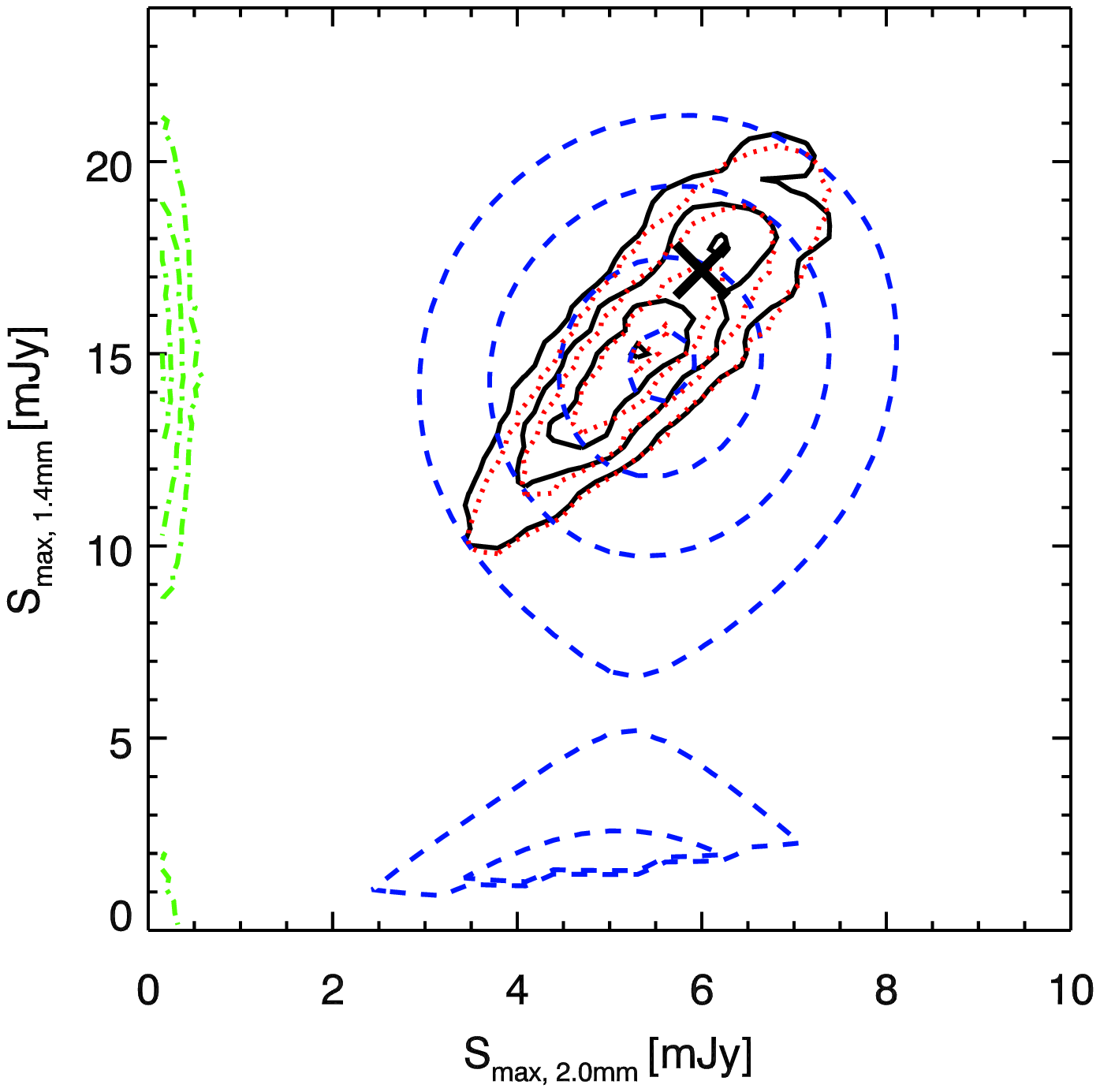, width=7cm} 
\end{center}
\caption{\scriptsize
Comparison of the true, underlying posterior two-band flux PDF 
$P(S_{\mathrm{max},1},S_{\mathrm{max},2}|S_{p,m,1},S_{p,m,2})$
(extracted from simulated observations) and calculated values for that
PDF for two possible pairs of measured $1.4$~mm and $2.0$~mm flux. 
{\bf Top Panel}:
$S_{p,m}=14.2$~mJy ($4.2 \sigma$) at $1.4$~mm, 
$S_{p,m}=7.0$~mJy ($4.9 \sigma$) at $2.0$~mm.
$2.0$~mm used as ``detection band" (see Sec.~\ref{sec:priorswap}).
{\bf Bottom Panel}:
$S_{p,m}=17.2$~mJy ($4.9 \sigma$) at $1.4$~mm, 
$S_{p,m}=6.0$~mJy ($4.1 \sigma$) at $2.0$~mm.
$1.4$~mm used as ``detection band".
Symbols and contours are as follows:
{\it Black X}:  Value of measured flux 
$S_{p,m}$ in the two bands.
{\it Black (solid) contours}:  true, underlying posterior two-band flux PDF,
extracted from the simulated observations 
as described in Sec.~\ref{sec:simstwoband}. 
(These contours have been smoothed slightly.)
{\it Blue (dashed) contours}:  posterior PDF calculated 
using the multi-band source flux estimation procedure described 
in Sec.~\ref{sec:multiband} with a flat prior on 
spectral index $\alpha$ with cutoff values of $-3$ and $5$.  
{\it Red (dotted) contours}:  similar to blue (dashed) contours, but
with a prior on $\alpha$ equal to the spectral index 
distribution used in the input to the simulated observations 
($\alpha = 2.7 \pm 0.3$).  
{\it Green (dash-dot) contours}:  posterior PDF obtained using 
the single-band procedure described in Sec.~\ref{sec:oneband} 
independently in both bands, ignoring the correlations between the
priors in the two bands.  All posterior PDF calculations use the 
\citet{negrello07} source counts model to construct the flux prior.
All contours are drawn at $0.5 \sigma$
intervals.
\label{fig:psmax_smeas_twoband}}
\end{figure}

Analogous to Fig.~\ref{fig:psmax_smeas} for the single-band case, 
Fig.~\ref{fig:psmax_smeas_twoband} shows comparisons between calculated
versions of the two-dimensional
posterior distribution $P(S_{\mathrm{max},1},S_{\mathrm{max},2}|S_{p,m,1},S_{p,m,2})$
and the true values of those distributions (extracted from simulated observations).
The simulated observations were performed by populating fake skies at two 
observing wavelengths ($1.4$~mm and $2.0$~mm) with a 
single population of sources with an underlying Gaussian distribution of 
spectral indices $\alpha = 2.7 \pm 0.3$, similar to the spectral index distribution
for high-redshift, dusty galaxies derived in \citet{knox04}.  The number counts as a function of flux
for the source population come from the \citet{negrello07} dusty galaxy counts at 
$850 \mu$m; this model was also used to construct the source flux prior.  
Noise was added to the fake skies in each band at a level similar 
to that in the corresponding SPT bands in  V09.  
To avoid confusing effects of slight misestimations of priors 
due to beam and filtering with fundamental issues in the two-band implementation, this 
set of simulated observations involved no spatial filtering beyond binning into 
$1$-arcmin pixels.  Posterior two-band flux PDFs were extracted from the simulated
observations as in the single-band case, namely by finding the brightest 
source associated with each detection in the simulated maps and constructing 
a two-dimensional histogram of $\{ S_{\mathrm{max},1},S_{\mathrm{max},2} \}$
for every pair of measured flux values.

True and calculated values of the posterior PDF are shown in 
Fig.~\ref{fig:psmax_smeas_twoband} for 
two values of measured two-band flux.  These raw flux values at $1.4$~mm 
and $2.0$~mm ---
$\{14.2,7.0\}$~mJy in the top panel of  
Fig.~\ref{fig:psmax_smeas_twoband} and 
$\{17.2,6.0\}$~mJy in the bottom panel of Fig.~\ref{fig:psmax_smeas_twoband} ---
correspond to detection significances of 
$\{4.2,4.9\}$ and
$\{4.9,4.1\}$.
These values were chosen to illustrate the importance of using the 
full two-band information as opposed to calculating each band's posterior 
flux PDF individually and ignoring correlations in the two bands' prior information.

Three different calculated values of the posterior are shown in each panel.
Two versions of the calculated posterior use the two-band implementation described
in Sec.~\ref{sec:multiband}: one using a flat prior on $\alpha$ ($-3 \le \alpha \le 5$) 
and one using a prior on $\alpha$ that is equal to the true underlying $\alpha$ 
distribution (a Gaussian with $\bar{\alpha}=2.7$ and $\sigma_\alpha=0.3$).
In each of these cases, the band in which the source was detected more 
significantly was used as the detection band.
The last version of the two-band posterior flux PDF shown in 
Fig.~\ref{fig:psmax_smeas_twoband} is the product of the individual single-band
posterior PDFs, each calculated using the procedure outlined in Sec.~\ref{sec:oneband} and
assuming no correlations between the prior information in each band.

The large amount of information in Fig.~\ref{fig:psmax_smeas_twoband} can
be boiled down to three main points:
\begin{enumerate}
\item If one had perfect prior information on $dN/dS$ and $\alpha$, one could 
calculate the posterior two-band flux PDF for every source perfectly.
\item With far less restrictive priors on the $\alpha$ distribution, one can make an
estimate of the posterior two-band flux PDF for every source that has no strong bias
but has somewhat less constraining power.
\item Calculating the posterior two-band flux PDF by de-boosting each source 
individually and assuming no correlation between the priors in each band can result
in highly biased posterior distributions.  The posterior flux estimate in the band in 
which the source is detected strongly is reasonable, but the flux estimate in the 
other band is de-boosted to the confusion limit.  The spectral index inferred from 
this calculation is actually a worse estimate of the true index than using the measured
flux in each band uncorrected for boosting (as shown by the black crosses in 
Fig.~\ref{fig:psmax_smeas_twoband}).  This highlights the importance of accounting
for the correlations in the prior information used to de-boost multi-band fluxes, particularly
when dealing with a source population for which the number counts are a steeply 
falling function of flux.
\end{enumerate}

\subsection{Classification of sources through their spectral behavior}
\label{sec:class}

\begin{figure}[h]
\begin{center}
\epsfig{file=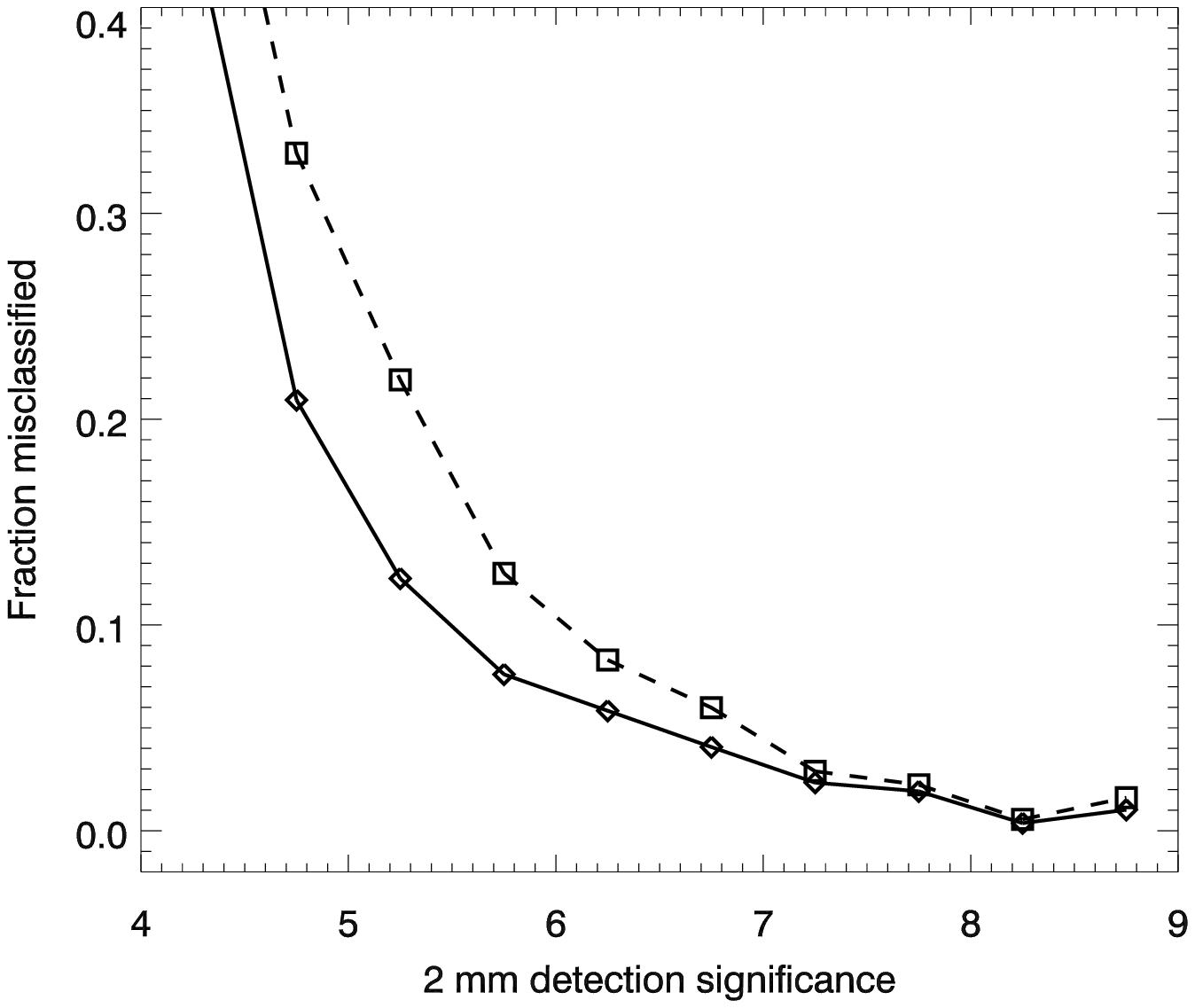,width=8cm} 
\epsfig{file=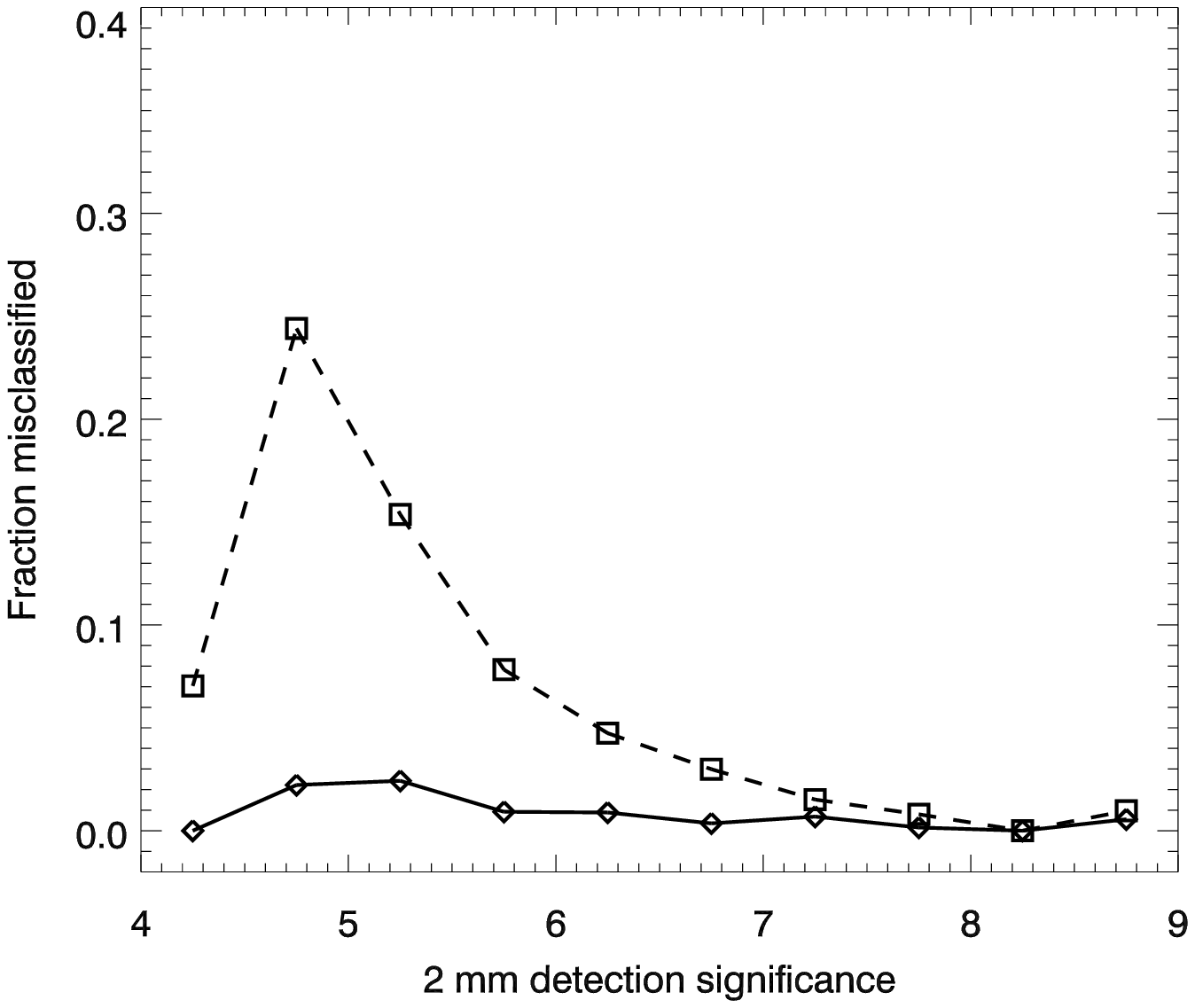,width=8cm} 
\end{center}
\caption{\scriptsize
Fraction of sources in simulated observations which would be  
misclassified by a posterior spectral index PDF criterion
which divides sources into synchrotron- 
or dust-dominated populations based on the value of
$P(\alpha \ge 1.5)$.  See Sec.~\ref{sec:class} for details of the 
simulated observation.  Diamonds and solid lines show the 
misclassification fraction obtained using the two-band posterior
flux estimation described in Sec.~\ref{sec:multiband};
boxes and dashed lines show the misclassification fraction 
obtained using the single-band procedure described in Sec.~\ref{sec:oneband} 
independently in both bands and ignoring the correlations in the
two bands' priors.
{\bf Top Panel}:  fraction of all sources misclassified as a function 
of $2.0$~mm detection significance.
{\bf Bottom Panel}:  fraction of sources misclassified which had 
$P(\alpha \ge 1.5) > 0.9$ or $P(\alpha \ge 1.5) < 0.1$, also as a function 
of $2.0$~mm detection significance.  The feature at $\sim 4.7 \sigma$
in the dashed curve in the bottom panel is an artifact of the 
specific noise levels chosen for the two bands (for details, see
Sec.~\ref{sec:class}).
\label{fig:misclass}}
\end{figure}

One key use of spectral information is to separate sources into 
different populations.  
For example, V09 use the posterior probability distribution of 
$\alpha$ for every detected source to classify it as either synchrotron- or 
dust-dominated. This allows the source counts in each population 
to be compared to models of that population and 
to counts at other wavelengths where that population is 
dominant.  Any bias in the posterior two-band flux 
PDF of a source could cause sources to be misclassified, leading 
to a bias in the estimation of source counts for both populations.

We have used simulated observations very similar to those used 
in Sec.~\ref{sec:simstwoband} to investigate how often sources are 
misclassified using the two-band posterior flux PDF estimated with 
and without accounting for correlations in the prior information 
between bands.  
The only difference in this set of 
simulated observations and those of Sec.~\ref{sec:simstwoband}
is that there are two populations used 
to create the fake skies that are observed by our instrument.  The 
two populations are the dust-dominated population used in 
Sec.~\ref{sec:simstwoband} plus a synchrotron-dominated 
population with source counts as in \citet{dezotti05} and a Gaussian spectral
index distribution with $\alpha = -0.7 \pm 0.5$, roughly consistent
with the spectral behavior of the brightest synchrotron-dominated
sources in  V09.  

For each detection in the simulated maps, 
the brightest source associated with that detection was identified, 
and the posterior spectral index PDF for the source was calculated
by initially calculating the two-dimensional posterior PDF for flux 
in one band and $\alpha$ and then marginalizing over the flux 
variable to create a one-dimensional spectral index PDF.  For this 
exercise, we only use the broad, flat $\alpha$ prior from 
Sec.~\ref{sec:simstwoband} ($-3 \le \alpha \le 5$).  The 
posterior $\alpha$ PDF was then compared to the true spectral 
index of the brightest source associated with that detection.  
We classified detections in the simulated maps as 
synchrotron-dominated if $P(\alpha \ge 1.5) < 0.5$
and dust-dominated if $P(\alpha \ge 1.5) > 0.5$; similarly, we classified 
the brightest source associated with the detection as synchrotron-
or dust-dominated according to whether its true spectral index was 
greater than or less than $1.5$.\footnote{The threshold
value of $\alpha=1.5$ was chosen to lie roughly at the minimum of 
the two-population spectral index histogram (for sources above the 
detection threshold)  The results in this section are insensitive to 
moving this threshold value by $\pm 0.5$.} 
The posterior
$\alpha$ PDF for each detection was also estimated by 
calculating the posterior flux PDF in each band independently
using the procedure outlined in Sec.~\ref{sec:oneband} --- ignoring any 
correlations between the prior information in the two bands --- 
and combining the 
two flux PDFs to create a PDF for $\alpha$.  This $\alpha$ PDF
was used to classify sources similarly to the two-band $\alpha$
posterior.

The fraction of sources misclassified (labeled as synchrotron-dominated
using the posterior $\alpha$ PDF when the brightest associated source 
was in fact dust-dominated, or vice-versa) in each case is shown
as a function of single-band detection significance
in Fig.~\ref{fig:misclass}.  The two-band implementation --- even with 
the weak prior on $\alpha$ --- shows a clear improvement in 
misclassification fraction over combining 
independent single-band PDFs at all significance levels up to 
$7 \sigma$.  Particularly striking is the difference in 
``high-confidence" misclassifications --- instances in which 
the classification based on the posterior $\alpha$ PDF was at 
the $90\%$ confidence level or greater, but was wrong.  As 
shown in the bottom panel of Fig.~\ref{fig:misclass}, the $\alpha$
estimation based on independent single-band PDFs has up 
to a $25\%$ rate of high-confidence misclassifications, but 
rate for the two-band implementation is effectively zero.  

The turnover at $\sim 4.7 \sigma$ in the rate of high-confidence
misclassifications using the single-band information and ignoring
correlations (Fig.~\ref{fig:misclass}, bottom panel, dashed curve)
is due to the relative noise level in the two bands.  Because the 
noise at $1.4$~mm is more than twice that at $2.0$~mm, there are 
many sources that are intrinsically dust-dominated which are 
nevertheless detected more significantly at $2.0$~mm.  If such 
a source is detected above $4.5 \sigma$ at $2.0$~mm but below
$4.5 \sigma$ at $1.4$~mm, the posterior flux PDF for that source 
will be centered near the raw, detected flux at $2.0$~mm but 
de-boosted to the confusion limit at $1.4$~mm.  This results in 
a robust ``measurement'' of a negative spectral index for this 
source and, hence, a high-confidence misclassification.  If the 
source is detected below $4.5 \sigma$ in both bands, there is 
effectively no constraint on $\alpha$ using this procedure, so 
the source may be misclassified, but not at high confidence.

\subsection{More than two bands}

The two-band formalism laid out in Sec.~\ref{sec:multiband} is 
sufficient for the V09 analysis of two-band SPT data.  However,
data in three or more bands at mm/submm wavelengths and mJy 
flux levels have been or are currently being collected 
by the Balloon-borne Large-Aperture Submillimeter 
Telescope\footnote{{\tt http://www.blastexperiment.info}}
 \citep[BLAST,][]{devlin09}, the SPT, and the Atacama Cosmology 
Telescope\footnote{{\tt http://www.physics.princeton.edu/act/}}
\citep[ACT,][]{fowler07}.  Furthermore, the recent launch of 
Herschel\footnote{{\tt http://www.esa.int/science/herschel}} 
and Planck\footnote{{\tt http://www.esa.int/SPECIALS/Planck/}}
means that we will soon have simultaneous measurements of 
mm/submm sources in as many as seven bands (depending
on where you choose to define the limits of the mm/submm 
spectral region).

Fortunately, the two-band formalism laid out in Sec.~\ref{sec:multiband} is easily
extended to more than two bands, although the calculation necessarily
becomes more complex.  For the case in which the Gaussian 
approximation holds and each source's spectral behavior can 
be described by a single power-law index across all bands, then the 
multi-band calculation is a trivial extension of Eqns.~\ref{eqn:psmax_alpha} - 
\ref{eqn:residvec}.  A first step in relaxing the assumption of a single 
spectral index for each source would be allowing a break in the 
spectrum such that each source would have a single spectral index 
between any two bands.  The spectral index prior would then be a 
function of $N_\mathrm{bands} - 1$ variables 
$\mbox {\boldmath $\alpha$} = \{ \alpha_{12}, \alpha_{13}, ..., \alpha_{1 N} \}$.
These variables would necessarily be highly correlated, so we would 
require the full $(N-1)$-dimensional prior.  A full
treatment of the $N$-band version of the method, including tests on simulated 
observations, will be the subject of future work.
 
\section{Conclusions}\label{sec:conclusions}
We have constructed a method for reliable, minimally biased estimation of 
single-band and multi-band properties of individual sources from noisy
data.  We find that proper treatment of 
correlated prior information in the multi-band version of the 
method is crucial to avoid significant biases in estimates of 
multi-band fluxes and spectral behavior.
The single- and multi-band implementations
of the method have been verified through  
simulated observations of mm data, and the two-band implementation
has been used to estimate source fluxes and spectral behavior 
in SPT data \citep{vieira09}.  This method, or an 
extension thereof to more than two bands, is directly applicable 
to source analyses for most current and upcoming 
mm/submm experiments, including BLAST, ACT, Planck, and 
Herschel and should also be applicable to data taken at
other wavelengths.

\acknowledgments
This work was supported by the National Science Foundation through 
grants ANT-0638937 and ANT-0130612 and the NSF Physics Frontier 
Center grant PHY-0114422 to the Kavli Institute of Cosmological Physics 
(KICP) at the University of Chicago.  ERS acknowledges support from a KICP
fellowship.  The authors would like to thank members 
of the SPT collaboration, Jason Austermann, and the anonymous referee 
for useful discussions and helpful suggestions.

\clearpage

\bibliography{../../spt_smg,../../../../BIBTEX/spt.bib}

\end{document}